\documentclass[twocolumn,secnumarabic,amssymb, nobibnotes, aps, prl,superscriptaddress]{revtex4-1}

\usepackage{graphicx}% Include figure files
\usepackage{dcolumn}% Align table columns on decimal point
\usepackage{bm}% bold math
\usepackage{hyperref}% add hypertext capabilities
\usepackage{natbib}

\begin{document}
	
	\preprint{APS/123-QED????????????}
	
	\title{Symmetry Protected Topological Metals}
	
	\author{Xuzhe Ying}
	\affiliation{School of Physics and Astronomy, University of Minnesota, Minneapolis, MN 55455, USA}
	
	\author{Alex Kamenev}
	\affiliation{School of Physics and Astronomy, University of Minnesota, Minneapolis, MN 55455, USA}
		\affiliation{William I. Fine Theoretical Physics Institute,  University of Minnesota,
Minneapolis, MN 55455, USA}


	
	
	\begin{abstract}
		We show that sharply defined topological quantum phase transitions are not limited to states of matter with gapped electronic spectra. Such transitions may also occur between two gapless metallic states both with extended Fermi surfaces. The transition is characterized by a discontinuous, but not quantized, jump in an off-diagonal transport coefficient. Its sharpness is 
		protected by a symmetry, such as e.g. particle-hole, which remains unbroken across the transition. We present 
		a simple model of this phenomenon, based on 2D $p+ip$ superconductor with an applied supercurrent, and discuss its geometrical interpretation.    
		%\begin{description}
		%\end{description}
	\end{abstract}
	
	\pacs{1111}% PACS, the Physics and Astronomy
	\maketitle

The advent of topological insulators and semimetals \cite{bernevig2013topological,shen2012topological,RevModPhys.82.3045,RevModPhys.83.1057,RevModPhys.88.035005} brought the realization that 
states of matter may be distinguished by subtle topological indexes. The very existence of such indexes 
primarily relies  on symmetries of the system, rather than its specific Hamiltonian \cite{RevModPhys.88.035005,ryu2010topological,wigner1958distribution,dyson1962threefold,PhysRevB.78.195125}. States with different topological indexes are separated by sharp quantum phase transitions (QPT), 
which are often associated with quantized jumps of certain transport coefficients (such as e.g. Hall conductance in 
integer quantum Hall effect\cite{prange1987quantum}).  Most of the current literature focuses on gapped phases (insulators or superconductors).
The exceptions are Weyl semimetals \cite{hosur2013recent,RevModPhys.90.015001}, with vanishing density of states at the Fermi level, 
and Anderson insulators \cite{PhysRevLett.102.136806,groth2009theory}, where all states away from QPT are strongly localized.  
	
The goal of this letter is to point out that the topological properties are not limited to  the gapped states of matter, or Weyl semimetals.  Instead,  they may persist well into a gapless metallic state with a finite density of delocalized states. Consequently, 
there are sharp  QPT's between topologically distinct metallic phases. One may dub 
them {\em topological metals} (TM), to distinguish from ordinary metals.  Across  QPT between TM and a metal a physical observable, associated with the topological index (e.g. an off-diagonal conductivity), exhibits a discontinuous jump. In contrast to topological QPT between two gapped phases,  such a jump is, in general,  {\em not} quantized. The topological QPT in metals should be necessarily protected by some symmetry. In the absence of any symmetry the metallic QPT  gives way to a smooth crossover, invalidating the sharp designation of TM phase.

\begin{figure}[t]
		\centering
		\includegraphics[scale=0.6]{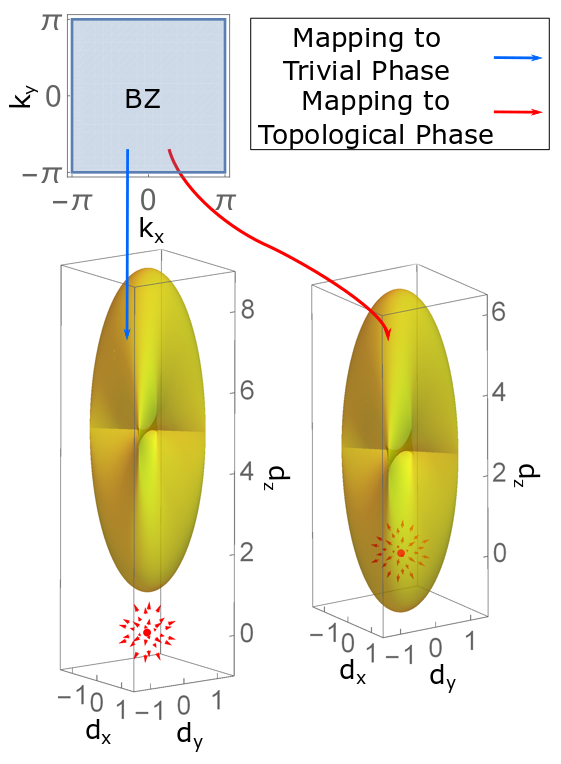}
		\caption{Brillouin zone mapping onto closed $\boldsymbol{d}(\boldsymbol{k})$ surface in 3D 
		$\boldsymbol{d}$-space for $t=1$, $\Delta=0.5$. The monopole, located at the origin $\boldsymbol{d}=0$, is shown in red. Trivial phase is shown for  $\mu=-5$, topological phase for $\mu=-2.5$. The Berry flux through the surface 
		is zero in the trivial phase and quantized in units of the monopole charge in the  topological phase.
		}
		\label{fig:MapBZ}
	\end{figure}

To illustrate these ideas we shall use a two-dimensional (2D) example, which belongs to the symmetry class D \cite{PhysRevB.55.1142,PhysRevB.78.195125,RevModPhys.82.3045,RevModPhys.83.1057,RevModPhys.88.035005}. This class is realized, for example, by $p+ip$ superconductors \cite{kallin2012chiral}, which break time reversal symmetry and the only protected symmetry is the particle-hole one. The latter is encoded within the Nambu structure of the corresponding  Bogoliubov-de Gennes (BdG) Hamiltonian, $H_{BdG}(\boldsymbol{k})$, where $\boldsymbol{k}$ is a quasi momentum in a 2D Brillouin zone (BZ), as \cite{RevModPhys.88.035005,RevModPhys.82.3045} 
\begin{equation}
					\label{Eq:symmetry} 
P^{-1}H_{BdG}(\boldsymbol{k})P=-H_{BdG}(-\boldsymbol{k}).					
\end{equation}
Here  $P=\sigma_x K$, where $K$ is complex conjugation operator and $\sigma_x$ is the Pauli matrix in Nambu space with the basis  $\Psi_{\boldsymbol{k}}=(c_{\boldsymbol{k}},\ c_{-\boldsymbol{k}}^{\dagger})^T$. A generic Hamiltonian has a form 
\begin{equation}
	H_{BdG}(\boldsymbol{k})=d_0(\boldsymbol{k})+\boldsymbol{d}(\boldsymbol{k})\cdot\boldsymbol{\sigma}
													\label{Eq:HG}
	\end{equation}
where $d_0(\boldsymbol{k})$ and $\boldsymbol{d}(\boldsymbol{k})=(d_x,d_y,d_z)$ are functions of momentum. The particle-hole symmetry, Eq.~(\ref{Eq:symmetry}), restricts $d_{0,x,y}(\boldsymbol{k})$  to be odd, while $d_z(\boldsymbol{k})$ even,  under $\boldsymbol{k}\leftrightarrow-\boldsymbol{k}$. The spectrum consists of two bands with energies  
\begin{equation}
							\label{eq:spectrum} 
\epsilon_{\boldsymbol{k}}^{(\pm)}=d_0(\boldsymbol{k})\pm\sqrt{d_x^2(\boldsymbol{k})+d_y^2(\boldsymbol{k})+d_z^2(\boldsymbol{k})},
\end{equation}
which may only touch when $\boldsymbol{d}=0$. In the simplest example of the square lattice \cite{read2000paired,bernevig2013topological}: $d_0=0$,  $d_x=-2\Delta\sin k_y$, $d_y=-2\Delta\sin k_x$ and $d_z=-2t\cos k_x-2t\cos k_y-\mu$ where $t$, $\mu$ and $\Delta$ are hopping parameter, chemical potential and p-wave pairing amplitude, correspondingly.  The  spectrum (\ref{eq:spectrum})  is fully gapped everywhere away from the topological QPT. The later takes place at  $\mu=\mp 4t$ and results in a gapless point at $\boldsymbol{k}=(0,0)$, or $(\pi,\pi)$, correspondingly.

\begin{figure}[t]
		\centering
		\includegraphics[scale=0.6]{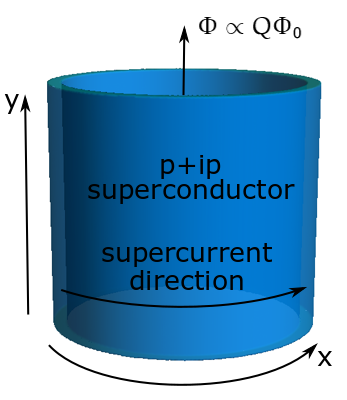}
		\caption{Schematic geometry of the system discussed in the text.}
		\label{fig:SysSpec}
	\end{figure}

The topological properties stem from the homotopy group $\mathbf{Z}$~\cite{RevModPhys.82.3045,RevModPhys.83.1057,RevModPhys.88.035005} associated with the mapping of the 2D BZ (torus) onto the 3D space spanned by the vector $\boldsymbol{d}$. (Notice that $d_0(\boldsymbol{k})$  component, being commutative with the Hamiltonian, is not related to the topology; it may be important however in assigning occupation numbers to  states  with momentum $\boldsymbol{k}$.) The image of BZ, 
$\boldsymbol{k}\in BZ$, is a closed 2D surface $\boldsymbol{d}(\boldsymbol{k})$ in the 3D    $\boldsymbol{d}$-space,  with an integer $\mathbf{Z}$ wrapping around the gapless point $\boldsymbol{d}=0$. The topological QPT, associated with the 
change of the integer wrapping number, occurs if the gapless point  $\boldsymbol{d}=0$ lies on the BZ image (in our example this only happens at $\mu=\mp 4t$), see Fig.~\ref{fig:MapBZ}. In the cylindrical geometry of Fig.~\ref{fig:SysSpec} the topological index counts a number of  gapless chiral modes localized near the two edges of the cylinder.

The physical  quantity, sensitive  to the  topological index,  is the intrinsic (anomalous) Hall conductance. In the case of the superconductor the object of interest is the {\em thermal} Hall conductance $\sigma_{xy}^\mathrm{int}$, given by the ratio of the thermal current in the $x$-direction  to the temperature gradient applied in the $y$-direction, Fig.~\ref{fig:SysSpec}.  
It originates from the anomalous velocity of   Bloch electrons due to the Berry curvature term  
\cite{PhysRevB.59.14915,PhysRevLett.93.206602} in the semiclassical equations of motion.  
According to Kubo-St\u reda formula \cite{streda1982theory,read2000paired}, the  
anomalous thermal Hall conductance (in unit of $\frac{\pi k_B^2}{12\hbar}T$) is given by the integrated Berry curvature:
	\begin{equation}
	\sigma_{xy}^\mathrm{int}= \sum_n\int_{BZ} \frac{d^2k}{(2\pi)^2}\, f\!\left(\epsilon^{(n)}_{\boldsymbol{k}}\right)\Omega_z^{(n)}(\boldsymbol{k}),
								\label{Eq:BerryPhase}
	\end{equation}
where  $\Omega_z^{(n)}(\boldsymbol{k})$ is the z-component of the Berry curvature, defined as the momentum space curl of the Berry connection $\boldsymbol{\Omega}^{(n)}(\boldsymbol{k})=\nabla_{\boldsymbol{k}}\times\boldsymbol{\mathcal{A}}^{(n)}(\boldsymbol{k})$ and $\boldsymbol{\mathcal{A}}^{(n)}(\boldsymbol{k})=\langle u^{(n)}(\boldsymbol{k})|i\nabla_{\boldsymbol{k}}|u^{(n)}(\boldsymbol{k})\rangle$. Here  $|u^{(n)}(\boldsymbol{k})\rangle$ is a Bloch wavefunction in the band $n$ and	$f(\epsilon_{\boldsymbol{k}}^{(n)})$ is the Fermi function.

For a fully gapped system at a temperature $T$ much less than the gap Eq.~(\ref{Eq:BerryPhase}) leads to a quantized anomalous conductance.  For example, the gapped two-band model, described by the Hamiltonian (\ref{Eq:HG}), 
results in 
	\begin{equation}
	                                       \label{Eq:monopole} 
\sigma_{xy}^\mathrm{int}	=\int_{BZ}\frac{d^2k}{(2\pi)^2}(\partial_{k_x}\boldsymbol{d}\times\partial_{k_y}\boldsymbol{d})\cdot \frac{\boldsymbol{d}}{2|\boldsymbol{d}|^3}\,.
	\end{equation}
This expression may be viewed as a flux of a monopole, located at $\boldsymbol{d}=0$, through the closed surface 
$\boldsymbol{d}(\boldsymbol{k})$, Fig.~\ref{fig:MapBZ}. Indeed, $d^2k(\partial_{k_x}\boldsymbol{d}\times\partial_{k_y}\boldsymbol{d})$ is the area element of the surface, while $\frac{\boldsymbol{d}}{2|\boldsymbol{d}|^3}$ is the field strength of the monopole with the unit ``charge''. Due to Gauss's law, such a flux is quantized and proportional to the integer
 wrapping number $\mathbf{Z}$ of the BZ image $\boldsymbol{d}(\boldsymbol{k})$ around  $\boldsymbol{d}=0$. This is the essence of the familiar conductance quantization in topological insulators  \cite{thouless1982quantized,laughlin1981quantized,bernevig2013topological,shen2012topological}.

\begin{figure*}
		\centering
		\includegraphics[scale=0.6]{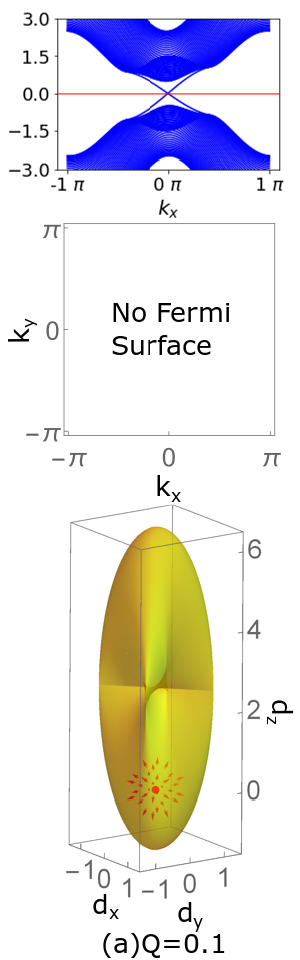}
		\includegraphics[scale=0.6]{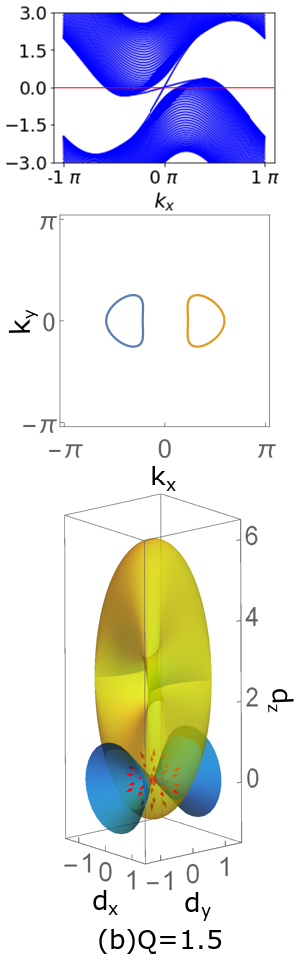}
		\includegraphics[scale=0.6]{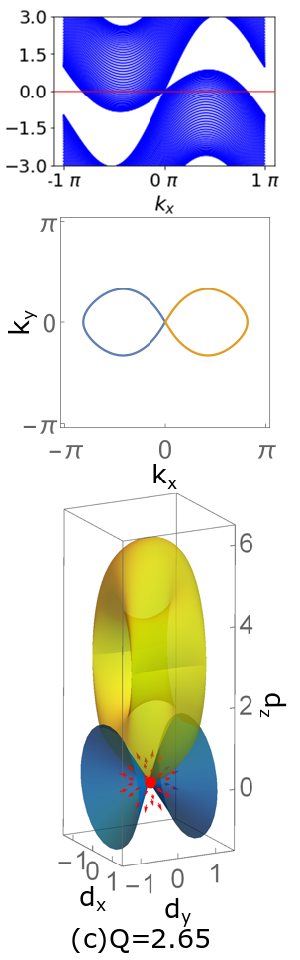}
		\includegraphics[scale=0.6]{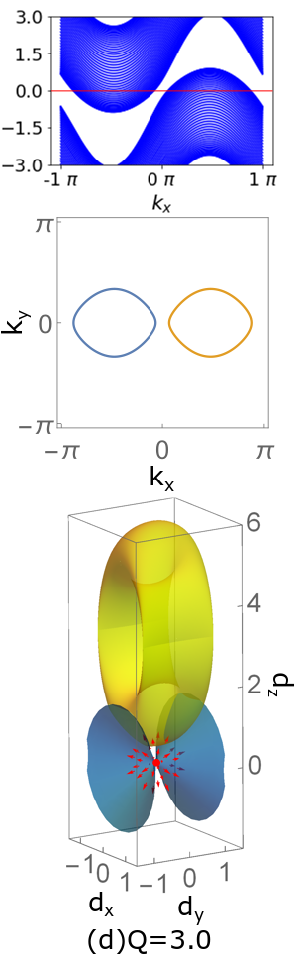}
		\caption{The three rows show spectra as function of $k_x$, BZ with the Fermi surfaces, 3D $\boldsymbol{d}$-space with the $\boldsymbol{d}(\boldsymbol{k})$ surface (gold), monopole at $\boldsymbol{d}=0$ (red) and Fermi double cone, Eq.~(\ref{Eq:cone}), (blue). The four columns correspond to different value of flux $Q$: (a) topological superconductor; (b) topological metal; (c) topological QPT; (d) ordinary metal. Figure~\ref{fig:PhaseDiagram} specifies other parameters.}
		\label{fig:BSFS}
	\end{figure*}

Let us now modify the model to bring it to the metallic state. The simplest way of doing it is to introduce a magnetic flux 
$\Phi$ through the cylinder of Fig.~\ref {fig:SysSpec}. The flux induces the supercurrent in the $x$-direction, breaking  the reflection symmetry, $\epsilon_{-\boldsymbol{k}}^{(n)}\neq \epsilon_{\boldsymbol{k}}^{(n)}$.  In the presence of the flux  the order parameter acquires a spatial dependence: $\Delta(x,y)=\Delta e^{iQx}$, where $Q=\Phi/(N\Phi_0)$ with $N$ number of lattice periods around the cylinder  \cite{footnoteProximity}. Upon a gauge transformation this leads to the Hamiltonian (\ref{Eq:HG}) with following parameters: 
	\begin{eqnarray}
		&&d_0=2t\sin k_x\sin Q/2,     \nonumber\\
		&&d_x=-2\Delta\sin k_y,\quad\quad  d_y=-2\Delta\sin k_x,               \label{Eq:d}\\
		&&d_z=-2t\cos k_x\cos Q/2-2t\cos k_y-\mu .   \nonumber 
	\end{eqnarray}
Notice that the particle-hole symmetry (\ref{Eq:symmetry}) and symmetry class D are still intact and so is the topological quantization of $\sigma_{xy}^\mathrm{int}$, as long as the spectrum (\ref{eq:spectrum}) remains fully gapped. This is indeed the case for sufficiently small flux $|Q|<Q_L=2\,\mathrm{arcsin}(\Delta/t)$. Figure~\ref{fig:BSFS}(a) shows spectrum for the cylindrical geometry of Fig.~\ref{fig:SysSpec}, which clearly exhibits chiral edge modes at $Q=.1<Q_L$. It also shows $\boldsymbol{d}(\boldsymbol{k})$ surface, which encloses the monopole at $\boldsymbol{d}=0$. 
	
At $|Q|=Q_L$ the system undergoes the Lifshitz transition\cite{lifshitz1960anomalies} into metallic state. This situation is depicted in Fig.~ \ref{fig:BSFS}(b), where one can clearly see two metallic bands: one is particle-like and the other is hole-like.  The corresponding Fermi surface consists of two disconnected closed curves in the 2D BZ. However, this is {\em not} yet the topological 
QPT, as one may notice by the presence of the chiral edge modes in the spectrum of Fig.~\ref{fig:BSFS}(b). 
Since the edge modes coexist now with the bulk states at the Fermi level, one does not expect a quantized thermal Hall conductance. Indeed, expression (\ref{Eq:monopole}) for the intrinsic conductance is still valid with the understanding that the integral runs  only over the $\boldsymbol{k}$-states with one occupied  band (at $T=0$). Thus 
the $\boldsymbol{d}(\boldsymbol{k})$ surface developes two holes -- the images of the 2D Fermi curves. 

\begin{figure}
		\centering
		\includegraphics[scale=0.6]{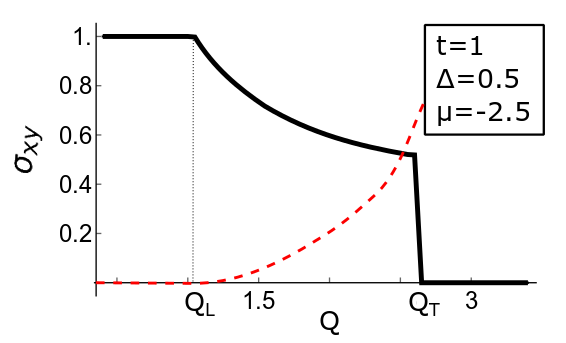}
		\caption{Thermal Hall conductance (in unit of $\frac{\pi k_B^2}{12\hbar}T$) vs. flux. $Q_L$ and $Q_T$ are location of Lifshitz and topological transitions.
		Solid line -- the intrinsic contribution, Eq.~(\ref{Eq:monopole}); dashed line -- (schematic) skew scattering contribution at $T\to 0$ limit. }
		\label{fig:BP}
	\end{figure}

\begin{figure}
		\centering
		\includegraphics[scale=0.6]{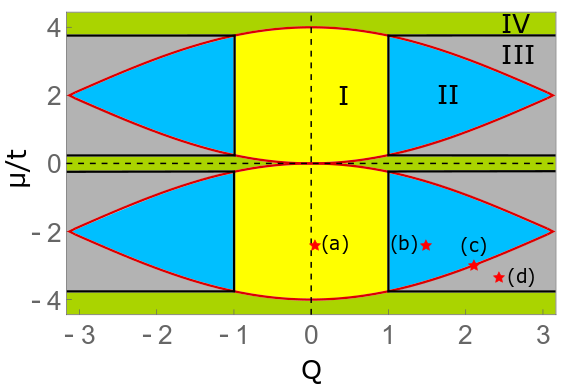}
		\caption{Phase diagram of the model Eqs.~(\ref{Eq:HG}), (\ref{Eq:d}) on chemical potential vs. flux plane.  I - topological insulator (superconductor); II -  topological metal; III - ordinary metal; IV - ordinary insulator. Solid red lines -- topological QPT,  solid black lines -- Lifshitz transitions. Red stars show parameters of columns (a-d) in Fig.~\ref{fig:BSFS}.}
		\label{fig:PhaseDiagram}
	\end{figure}

To understand it geometrically one may notice that $d_0=-\frac{t}{\Delta}\sin (Q/2)d_y$ and therefore the equations for the Fermi curves $\epsilon_{\boldsymbol{k}}^{(\pm)}=0$ acquire the form (cf. Eqs.~(\ref{eq:spectrum}), (\ref{Eq:d})): 
	\begin{equation}
						\label{Eq:cone}
	d_x^2 + \left[1-\frac{\sin^2 (Q/2)}{\sin^2 (Q_L/2)}\right]d_y^2+ d_z^2=0, 
	\label{cone}
	\end{equation}
where $\sin (Q_L/2) =\Delta/t$. For $|Q|>Q_L$ this condition spells the  {\em double cone} in the $\boldsymbol{d}$-space with the apex at the monopole  $\boldsymbol{d}=0$. The images of the Fermi curves are thus found as the intersections of the cone, Eq.~(\ref{Eq:cone}), with the closed surface $\boldsymbol{d}(\boldsymbol{k})$, Figs.~\ref{fig:BSFS}(b-d). The flux of the monopole, which contributes to  $\sigma_{xy}^\mathrm{int}$, Eq.~(\ref{Eq:monopole}), is therefore less than the quantized value by the amount of the flux channeled through the cone  (\ref{Eq:cone})  
\begin{equation}
						\label{Eq:sigma-Q}
\sigma_{xy}^\mathrm{int}(Q)=\sin\frac{Q_L}{2}/|\sin\frac{Q}{2}|, 
\end{equation}
where $|\sin (Q/2)|\geq \sin (Q_L/2)$, see Fig.~\ref{fig:BP}. 
The phase diagram of the system is schematically depicted in Fig.~\ref{fig:PhaseDiagram}. At the Lifshitz transition the 
system goes from the topological insulator (superconductor) phase to the TM phase. It is is characterized by the coexistence of the bulk states at the Fermi level with the chiral edge modes. 
The latter are responsible for the intrinsic contribution to the thermal Hall conductance, which is {\em not} quantized.      

It turns out that the Lifshitz transition is followed by another transition at $|Q|=Q_T(\mu)> Q_L$, where  $\cos (Q_T/2)=1-\frac{|\mu|}{2t}$.  This second transition separates two topologically distinct {\em metallic} phases. At the transition the two Fermi curves touch each other at the single point  $\boldsymbol{d}=0$, Fig.~\ref{fig:BSFS}(c). On the other side of the transition the two Fermi curves separate 
again,  Fig.~\ref{fig:BSFS}(d), and the edge states disappear. In the 3D $\boldsymbol{d}$-space  the apex of the cone (\ref{Eq:cone}) crosses the surface $\boldsymbol{d}(\boldsymbol{k})$ and the Berry flux of the monopole, Eq.~(\ref{Eq:monopole}), undergoes a discontinuous jump down to zero. The non-quantized hight of the jump is given by $\delta\sigma_{xy}^\mathrm{int}= \sin\frac{Q_L}{2}/\sin\frac{Q_T}{2}<1$, cf. Eq.~(\ref{Eq:sigma-Q}).  

The sharp topological QPT at $Q_T$ allows for unambiguous distinction between TM and the ordinary metal states. This sharp distinction is protected by the particle-hole symmetry, Eq.~(\ref{Eq:symmetry}). Indeed, since the surface  $\boldsymbol{d}(\boldsymbol{k})$ is punctured by the holes created by the Fermi curves, one could naively expect the Berry flux and $\sigma_{xy}^\mathrm{int}$ to evolve to zero in a smooth continuous way. This would be the case, if the gapless point $\boldsymbol{d}=0$ could move (as a function of some parameters) through one of those Fermi punctures. Such  scenario would invalidate the notion of the well defined TM phase. It is precluded  by the  particle-hole symmetry, Eq.~(\ref{Eq:symmetry}), which ensures that 
$\boldsymbol{d}=0$ point can't fall inside any of the Fermi curves, but can only simultaneously touch both of them. The double cone construction, Eq.~(\ref{Eq:cone}), is a geometric manifestation of the symmetry (\ref{Eq:symmetry}). It shows that 
the Berry flux through the punctured  $\boldsymbol{d}(\boldsymbol{k})$ surface must change discontinuously at the  topological QPT. 

Let us now briefly discuss the role of disorder. The latter has two distinct effects on the discussed phenomena. In the metallic phase (being treated beyond the Born approximation) it generates additional contribution to the thermal Hall conductance, known as the skew-scattering \cite{RevModPhys.82.1539,nagaosa2006anomalous,PhysRevB.75.045315,li2015anomalous,ado2015anomalous,PhysRevLett.118.027001},  Fig.~\ref{fig:BP}. Its specific value depends on the details of the disorder \cite{li2015anomalous,ado2015anomalous} and may exceed the intrinsic contribution, discussed here. The important observation is that the skew-scattering contribution, being a bulk phenomenon, is continuous across the topological phase transition  at $Q=Q_T$ \cite{YKToBePublished}.   
It therefore does not alter the discontinuity in $\sigma_{xy}$, but merely adds a smooth background.  

The second effect of the disorder is associated with the modification of the intrinsic contribution itself. We performed numerical 
simulations on small (way smaller than the localization length) lattices in the cylindrical geometry \cite{YKToBePublished}. It showed that for each disorder realization the discontinuity  $\sigma_{xy}^\mathrm{int}$ exists, though it's location and height fluctuate from one realization to another. In the thermodynamic limit, we expect  the Anderson localization to stabilize the topological transition \cite{khmelnitskii1983quantization,pruisken1990field,wei1986localization,bocquet2000disordered,PhysRevLett.112.206602}, in a way similar to the integer quantum Hall effect. However, such a transition separates now topologically distinct Anderson insulator, rather than metallic, phases. Though a full theory of such transition in 2D class D \cite{bocquet2000disordered} is still absent, it's likely that localization restores  quantization of  $\sigma_{xy}$.

To conclude, we have shown that the sharp definition of  the topological states may be extended onto a gapless metallic 
phase. An unbroken symmetry is required to enforce identity of such topological metal state. As an example we worked out 
class D, $p+ip$ superconductor subject to a super-current.  The TM phase, protected by the particle-hole symmetry,  appears in a certain finite range of the super-current densities.   It may be detected by a jump of the thermal Hall conductance, associated with the discontinuous  change of the Berry curvature flux.  

We are grateful to A. Andreev, I. Burmistrov, F. Burnell, A. Chudnovskiy, T. Gulden, M. Sammon and Z. Yang for useful discussions. 
Numerical simulations were aided  by the KWANT \cite{groth2014kwant} open source code.
This work was supported by NSF grant DMR-1608238.

	\bibliography{TMRef}{}

\begin{thebibliography}{37}
\bibitem [{foo()}]{footnoteProximity}%
  \BibitemOpen
  \href@noop {} {}\bibinfo {note} {We imagine $p+ip$ superconductivity as being
  induced by proximity to a strong $s$-wave superconductor. We thus do not
  consider self-consistency condition and dependence of the amplitude
  $|\Delta|$ on the flux $Q$}\BibitemShut {NoStop}%
\end{thebibliography}

\end{document}